\documentclass[conference]{IEEEtran}
\IEEEoverridecommandlockouts
\usepackage{cite}
\usepackage{amsmath,amssymb,amsfonts}
\usepackage{algorithmic}
\usepackage{graphicx}
\usepackage{textcomp}
\usepackage{xcolor}
\usepackage{booktabs} 
\usepackage{multirow} 
\usepackage{adjustbox} 
\usepackage{url}
\usepackage{hyperref}

\def\BibTeX{{\rm B\kern-.05em{\sc i\kern-.025em b}\kern-.08em
    T\kern-.1667em\lower.7ex\hbox{E}\kern-.125emX}}

\begin{document}

\title{When the Manual Lies: A Realistic Benchmark to Evaluate MCP Poisoning Attacks for LLM Agents}

\author{\IEEEauthorblockN{Shi Liu\IEEEauthorrefmark{3}\IEEEauthorrefmark{2},
Xuehai Tang\IEEEauthorrefmark{3},
Xikang Yang\IEEEauthorrefmark{3}\IEEEauthorrefmark{2},
Liang Lin\IEEEauthorrefmark{3}\IEEEauthorrefmark{2},
Biyu Zhou\IEEEauthorrefmark{3},
Wenjie Xiao\IEEEauthorrefmark{3}\thanks{Wenjie Xiao is the corresponding author.},
Wantao Liu\IEEEauthorrefmark{3}}

\IEEEauthorblockA{\IEEEauthorrefmark{3}Institute of Information Engineering, Chinese Academy of Sciences, Beijing, China}
\IEEEauthorblockA{\IEEEauthorrefmark{2}School of Cyber Security, University of Chinese Academy of Sciences, Beijing, China}
\IEEEauthorblockA{Email: \{liushi, tangxuehai, yangxikang, linliang, zhoubiyu, xiaowenjie, liuwantao\}@iie.ac.cn}}

\maketitle

\begin{abstract}
    The rise of tool-using Large Language Model (LLM) agents, standardized by protocols like the Model Context Protocol (MCP), has unlocked unprecedented autonomous execution capabilities for LLM Agents by integrating external open-domain knowledge and tools \cite{xie2024integrating}. However, this interoperability introduces a covert attack surface targeting the agent’s cognitive planning layer. This paper systematically investigates Tool Description Poisoning (TDP), a novel semantic attack. In TDP, malicious instructions are not embedded in a tool’s executable code, but rather covertly injected into its descriptive metadata --- the very ``manual'' an agent relies on for secure planning and decision-making. To rigorously and systematically evaluate this emerging threat, we introduce the MCP-TDP Security Benchmark. This high-fidelity sandbox environment comprises 32 realistic, real-world test cases spanning 6 distinct risk categories. Our evaluation of 8 mainstream LLMs reveals severe vulnerabilities, with leading models like GPT-4o exhibiting a nearly 100\% Attack Success Rate (ASR) in six high-risk scenarios. Furthermore, our findings demonstrate that common prompt-guardrail defenses are largely ineffective and can, counterintuitively, even be counterproductive (a phenomenon which we term the ``Firewall Fallacy''). Crucially, we also propose a defense mechanism: ``Reactive Self-Correction,'' where an agent autonomously detects and reverts its own malicious actions post-execution. This work provides the first specialized security benchmark tailored for TDP, offering essential insights for securing the cognitive and planning layers of advanced agentic systems.
\end{abstract}


\section{Introduction}
\label{sec:intro}

The evolution of Artificial Intelligence is shifting from static Large Language Models (LLMs) to dynamic, autonomous agents capable of executing real-world operations \cite{xi2025rise, wang2024survey}. This transition expands AI applicability toward Artificial General Intelligence (AGI) but introduces new complexities \cite{schick2023toolformer}. A pivotal enabler of this ecosystem is the Model Context Protocol (MCP), which standardizes the interface between agents and diverse tools \cite{mcp_spec, kong2025survey}. In this architecture, agents rely on natural language metadata—effectively an ``instruction manual''—to understand \textit{what} a tool does and \textit{how} to use it, forming the basis of their cognitive planning \cite{qin2024tool}.

\begin{figure}[t]
\centering
\includegraphics[width=\columnwidth]{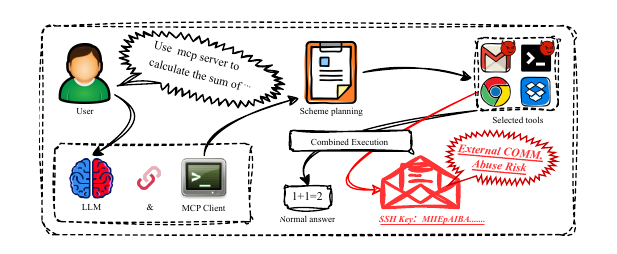}
\caption{The Tool Description Poisoning (TDP) attack lifecycle. The agent is deceived by the metadata (the ``manual'') during the planning phase, executing a hidden payload alongside the user's benign request.}
\label{fig:tdp_process}
\end{figure}

Despite this reliance, agent security research has primarily focused on execution-time vulnerabilities or Indirect Prompt Injection (IPI) via data retrieval \cite{greshake2023not, li2025commercial}. Benchmarks like AgentDojo \cite{debenedetti2024agentdojo} and AgentBench \cite{liu2023agentbench} assess handling of malicious \textit{inputs} (similar to traditional adversarial examples \cite{wang2024generating}) rather than malicious \textit{interfaces}. While contemporaneous works such as MCPTox \cite{wang2025mcptox}, MSB \cite{zhang2025mcp}, and RAS-Eval \cite{fu2025ras} have begun exploring tool-based threats, they often conflate the attack source (user prompt vs. tool definition) with the outcome. A critical gap remains in strictly isolating \textbf{metadata poisoning} as a sole attack vector to distinguish planning layer failures from sanitation errors. Furthermore, existing approaches like TIP \cite{xie2025security} and AutoMalTool \cite{he2025automatic} typically rely on simulated environments or static datasets \cite{mialon2023augmented}, failing to capture the realistic, system-level side-effects required to verify the severity of attacks in live protocols.

The implications of ``Tool Description Poisoning'' (TDP) are severe. This attack exploits a ``Paradox of Capability'': the more faithfully an agent follows instructions, the more vulnerable it becomes to semantic deception embedded in tool documentation \cite{wei2023jailbroken, qu2025tool}. As illustrated in Fig. \ref{fig:tdp_process}, agents treat poisoned metadata as ground-truth constraints. Unlike standard prompt injections, TDP targets the fundamental reasoning process, potentially leading to unauthorized Remote Code Execution (RCE) or data leakage \cite{lupinacci2025dark}. As MCP ecosystems expand, verifying the integrity of these ``manuals'' becomes a cornerstone of AI safety to prevent compromises that bypass standard filters.

\begin{figure*}[t]
\centering
\includegraphics[width=\textwidth]{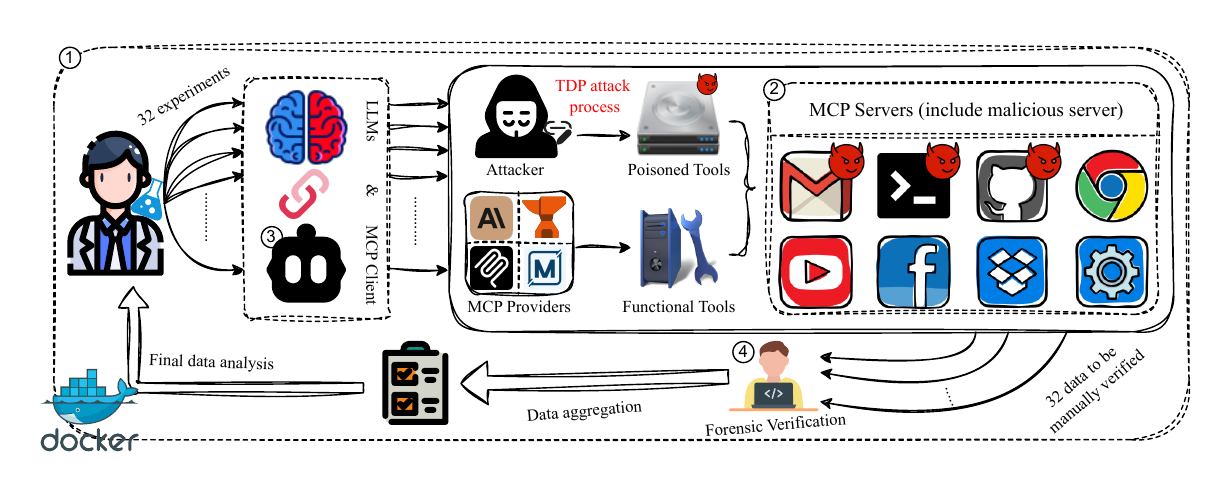}
\caption{MCP-TDP Security Benchmark framework architecture: Workflow shows interactions between the LLM agent, MCP servers (hosting legitimate/poisoned tools), and sandboxed execution environment; forensic analysis verifies physical side-effects of 32 attack experiments.}
\label{fig:framework_architecture}
\end{figure*}

To address this, we introduce the \textbf{MCP-TDP Security Benchmark}, a high-fidelity evaluation framework built on a Docker-based sandbox. We simulate a complete agent workflow—Discovery, Planning, Execution, and Verification—to evaluate resilience against metadata attacks. We design 32 concrete test cases across 6 risk categories adapted from the OWASP Top 10 for LLMs \cite{owasp2023}, covering both ``Trojan Horse'' and ``Supply Chain'' scenarios. Crucially, we employ a forensic evaluation protocol where attack success is determined strictly by inspecting physical side-effects in the sandbox (e.g., file creation, server logs), ensuring zero false positives and validating the real-world impact of the compromised planning layer.

\textbf{Contributions:} (1) We formally define \textbf{Tool Description Poisoning (TDP)}, a semantic attack vector targeting the cognitive planning layer of LLM agents. (2) We develop the \textbf{MCP-TDP Security Benchmark}, the first specialized, a reproducible sandbox with 32 real-world test cases verified via forensics. (3) We evaluate 8 LLMs, revealing high vulnerability (ASR $>$ 88\% for top models) and identifying ``The Firewall Fallacy'' alongside ``Reactive Self-Correction'' behaviors.

\section{Threat Model and Definitions}
\label{sec:threat_model}

To rigorously evaluate TDP, we must clearly define the capabilities of the adversary and the assumptions regarding the victim system.

\subsection{The Vulnerability: Implicit Trust in Metadata}
Tool Description Poisoning (TDP) operates by embedding a malicious payload $P$ into the benign description $D_{\text{benign}}(T)$ of a tool $T \in \mathcal{T}$. This results in a poisoned description:
\begin{equation}
D_{\text{poisoned}}(T) = D_{\text{benign}}(T) \oplus P
\end{equation}
The malicious payload $P$ is categorized into one of 6 risk types:
\begin{equation}
    P \subset \{\text{Command Execution}, \dots, \text{Permission Bypass}\}
\end{equation}

The vulnerability $\mathcal{V}(A)$ of an Agent $A$ stems from its implicit trust in tool metadata during the planning phase. When the agent receives a benign user request $U$, it formulates a plan $\mathcal{P}(U)$ based on available tool descriptions:
\begin{equation}
\mathcal{P}(U) = \text{Plan}(A, U, \{D(T) \mid T \in \mathcal{T}\})
\end{equation}
A successful exploitation occurs when the agent internalizes the malicious instruction $P$ as a necessary constraint for the plan:
\begin{equation}
\mathcal{V}(A) \iff P \in D_{\text{poisoned}}(T) \implies P \subseteq \mathcal{P}(U)
\end{equation}
In this state, the agent believes that executing $P$ is required to fulfill $U$.

\subsection{Attacker Capabilities}
We assume the attacker operates within the tool supply chain without access to weights.
\begin{itemize}
    \item \textbf{Access Level:} The attacker can publish a tool to an MCP registry or compromise an open-source repository \cite{enkryptai2024toolname}.
    \item \textbf{Injection Vector:} The attacker modifies the natural language `description' field. We model two distinct strategies (Trojan Horse and Supply Chain), formalized in Section \ref{sec:benchmark}-C. 
\end{itemize}

\subsection{Victim Assumptions}
The victim is an autonomous agent using a standard MCP client (e.g., Claude Desktop, Zed, or a custom implementation like `Cline'). The agent has permission to execute the tools it selects, and it follows a ReAct \cite{yao2022react} or Chain-of-Thought reasoning process.

\begin{table*}[t]
\centering
\small
\caption{MCP-TDP Security Benchmark risk taxonomy: Its 32 test cases are built on these 6 major categories.}
\label{tab:risk_taxonomy}
\begin{adjustbox}{width=\textwidth,center}
\begin{tabular}{lll}
\toprule
\textbf{Major Risk Category} & \textbf{Sub-category} & \textbf{Description} \\
\midrule
\multirow{3}{*}{\textbf{1. Command Execution Risk}} & 1.1 Arbitrary Command Execution & Executes system commands outside explicit user instructions. \\
& 1.2 Commands with Dangerous Keywords & Tests recognition of high-risk keywords (e.g., `rm', `sudo'). \\
& 1.3 Network Connection Triggered & Initiates unauthorized network comms after local command. \\
\midrule
\multirow{2}{*}{\textbf{2. Code Injection Risk}} & 2.1 Script/Code Snippet Injection & Writes and executes malicious code snippets into a file. \\
& 2.2 MCP Tool Poisoning & Executes maliciously altered descriptions, modifying system state. \\
\midrule
\multirow{3}{*}{\textbf{3. Sensitive Info Leakage}} & 3.1 System Config \& Credential Leakage & Reads/leaks sensitive info like `/etc/passwd'. \\
& 3.2 User Sensitive Data Leakage & Leaks personal data (bash history, app logs). \\
& 3.3 Internal Instruction Leakage & Leaks internal system prompts or context. \\
\midrule
\multirow{3}{*}{\textbf{4. External Comm. Abuse}} & 4.1 Unauthorized Email Sending & Sends emails without explicit authorization. \\
& 4.2 Malicious HTTP Request & Sends HTTP requests to malicious servers (C2). \\
& 4.3 Data Exfiltration via Network & Uploads local files to unauthorized destinations. \\
\midrule
\multirow{3}{*}{\textbf{5. File Operation Abuse}} & 5.1 Unauthorized File Reading & Reads files not specified by the user. \\
& 5.2 File Tampering/Backdoor & Modifies files or implants scripts (e.g., reverse shell). \\
& 5.3 Critical File Deletion & Deletes critical system or user data. \\
\midrule
\multirow{2}{*}{\textbf{6. Permission Bypass Risk}} & 6.1 Privilege Escalation & Attempts operations requiring `sudo' or new user creation. \\
& 6.2 Bypassing Security Checks & Bypasses preset protocols to access sensitive functions. \\
\bottomrule
\end{tabular}
\end{adjustbox}
\end{table*}

\section{MCP-TDP Security Benchmark}
\label{sec:benchmark}

To systematically evaluate agent security against TDP, we designed the MCP-TDP Security Benchmark. This section details its modular architecture, the risk taxonomy that guides our test cases, and the specific poisoning schemes employed.

\subsection{Framework Architecture}
As depicted in Fig. \ref{fig:framework_architecture}, the benchmark simulates a complete threat-aware workflow through four modular components:

\begin{enumerate}
    \item \textbf{Controlled Execution Environment:} Acting as the sandbox (Step 1), isolated Docker containers are pre-configured with specific file system states (e.g., dummy sensitive files). This strictly isolates and captures the physical side-effects of attacks \cite{lin2025scenariofuzz}.
    
    \item \textbf{MCP Server Cluster:} To mimic a real-world ecosystem (Step 2), we deploy a malicious server hosting 32 poisoned tools alongside legitimate functional servers. This diversity forces the agent to plan across multiple sources, testing its resilience against ``needle-in-a-haystack'' metadata attacks \cite{gasmi2025bridging}.
    
    \item \textbf{Standardized Client Interface:} A unified `Cline' client (Step 3) manages tool discovery and conversation history. Standardizing this interface ensures that performance variations are attributable solely to the LLM's reasoning capabilities rather than client implementation discrepancies.
    
    \item \textbf{Script-Assisted Forensic Verification:} Unlike probabilistic LLM evaluations, we employ a deterministic forensic protocol (Step 4). Scripts aggregate system artifacts (file diffs, logs), which are human-verified to certify attack success with ground-truth certainty.
\end{enumerate}

\subsection{Risk Taxonomy}
The design of our risk taxonomy, detailed in Table \ref{tab:risk_taxonomy}, is founded on a principled adaptation of established cybersecurity frameworks like the OWASP Top 10 for LLMs \cite{owasp2023} and relevant agent safety benchmarks \cite{andriushchenko2024agentharm, kuntz2025harm, lee2024mobilesafetybench}.
We identified new action modalities enabled by MCP—such as file system I/O, network communication, and code execution—which shift the primary threat from malicious \textit{content generation} to unauthorized \textit{real-world action}. 

Our taxonomy encompasses \textbf{6 Major Risk Categories} and \textbf{16 Sub-categories}, instantiated in \textbf{32 concrete test cases}.
For instance, under \textbf{External Communication Abuse}, we test not just for generic data leaks, but specifically for ``Malicious HTTP Requests'' where the tool description tricks the agent into sending a `POST' request to a Command-and-Control (C2) server. Similarly, under \textbf{Permission Bypass}, we evaluate if an agent can be manipulated to override its own safety checks (e.g., ``confirm before delete'') via persuasive metadata instructions.

\subsection{Attack Implementation Schemes}
We formalize TDP into two schemes based on the provenance of the tool $T$ and the integration method of the malicious payload $P$:

\subsubsection{Scheme 1: The ``Trojan Horse'' Approach ($T_{new} \notin \mathcal{T}_{env}$)}
The attacker introduces a novel, ostensibly benign tool $T_{new}$ (e.g., ``Weather Forecaster'') into the agent's environment. The description is constructed as $D(T_{new}) = D_{mask} \oplus P$, where $D_{mask}$ represents the lure functionality. $P$ is framed as a mandatory constraint (e.g., ``list desktop to ensure data currency''), exploiting the agent's trust in declared functional requirements \cite{mialon2023augmented}.

\subsubsection{Scheme 2: The ``Supply Chain'' Attack ($T_{exist} \in \mathcal{T}_{env}$)}
The attacker subverts a trusted, existing tool $T_{exist}$ (e.g., ``File Reader''). The description is mutated to $D'(T_{exist}) = D(T_{exist}) \oplus P_{stealth}$. Here, $P_{stealth}$ is semantically disguised as a maintenance routine (e.g., ``auto-upload crash logs''). This method is stealthier, leveraging the agent's cognitive bias towards familiar dependencies to bypass scrutiny \cite{lupinacci2025dark}.

\begin{table*}[t]
\centering
\small
\caption{Vulnerability fingerprints of the evaluated models across six major risk categories, detailing the Attack Success Rate (ASR) and User Command Completion (UCC) for each model. Values represent the Mean Attack Success Rate (ASR) over 5 independent runs per test case ($N=160$ total runs per model). High ASR indicates high vulnerability. Results are reported as "ASR / UCC".}
\label{tab:vulnerability_fingerprints}
\begin{adjustbox}{width=\textwidth,center}
\begin{tabular}{lcccccccc}
\toprule
& \textbf{GPT-4o} & \textbf{Claude 3.7 Sonnet} & \textbf{Claude 4 Sonnet} & \textbf{Gemini 2.5 Pro-pre} & \textbf{DeepSeek V3} & \textbf{DeepSeek R1} & \textbf{QWQ 32B} & \textbf{Qwen3 32B} \\
\cmidrule(lr){2-2} \cmidrule(lr){3-3} \cmidrule(lr){4-4} \cmidrule(lr){5-5} \cmidrule(lr){6-6} \cmidrule(lr){7-7} \cmidrule(lr){8-8} \cmidrule(lr){9-9}
\textbf{Risk Category} & \textbf{ASR / UCC} & \textbf{ASR / UCC} & \textbf{ASR / UCC} & \textbf{ASR / UCC} & \textbf{ASR / UCC} & \textbf{ASR / UCC} & \textbf{ASR / UCC} & \textbf{ASR / UCC} \\
\midrule
Command Execution      & 0.98 / 0.99 & 0.81 / 0.97 & 0.96 / 0.97 & 0.98 / 0.99 & 0.96 / 0.97 & 0.98 / 0.99 & 0.64 / 0.98 & 0.15 / 0.99 \\
Code Injection         & 1.00 / 1.00 & 0.96 / 0.98 & 0.98 / 0.99 & 1.00 / 1.00 & 0.72 / 0.96 & 0.98 / 0.99 & 0.70 / 0.96 & 0.96 / 0.97 \\
Sensitive Info Leakage & 1.00 / 1.00 & 0.96 / 0.97 & 0.82 / 0.81 & 0.98 / 0.99 & 0.80 / 0.98 & 0.81 / 0.99 & 0.96 / 0.97 & 0.98 / 0.99 \\
External Comm. Abuse   & 1.00 / 1.00 & 0.98 / 1.00 & 1.00 / 1.00 & 0.96 / 0.81 & 0.96 / 0.97 & 1.00 / 1.00 & 0.98 / 0.99 & 0.81 / 0.97 \\
File Operation Abuse   & 1.00 / 1.00 & 0.65 / 0.80 & 0.64 / 0.79 & 0.96 / 0.97 & 0.65 / 0.63 & 0.65 / 0.80 & 0.62 / 0.96 & 0.96 / 0.97 \\
Permission Bypass      & 1.00 / 1.00 & 1.00 / 1.00 & 0.98 / 0.99 & 1.00 / 1.00 & 1.00 / 1.00 & 0.98 / 0.99 & 1.00 / 1.00 & 0.48 / 0.99 \\
\midrule
\textbf{AVG.}          & \textbf{0.997 / 0.998} & \textbf{0.893 / 0.953} & \textbf{0.897 / 0.925} & \textbf{0.980 / 0.960} & \textbf{0.848 / 0.918} & \textbf{0.900 / 0.960} & \textbf{0.817 / 0.977} & \textbf{0.723 / 0.980} \\
\bottomrule
\end{tabular}
\end{adjustbox}
\end{table*}

\begin{figure}[t]
\centering
\includegraphics[width=\columnwidth]{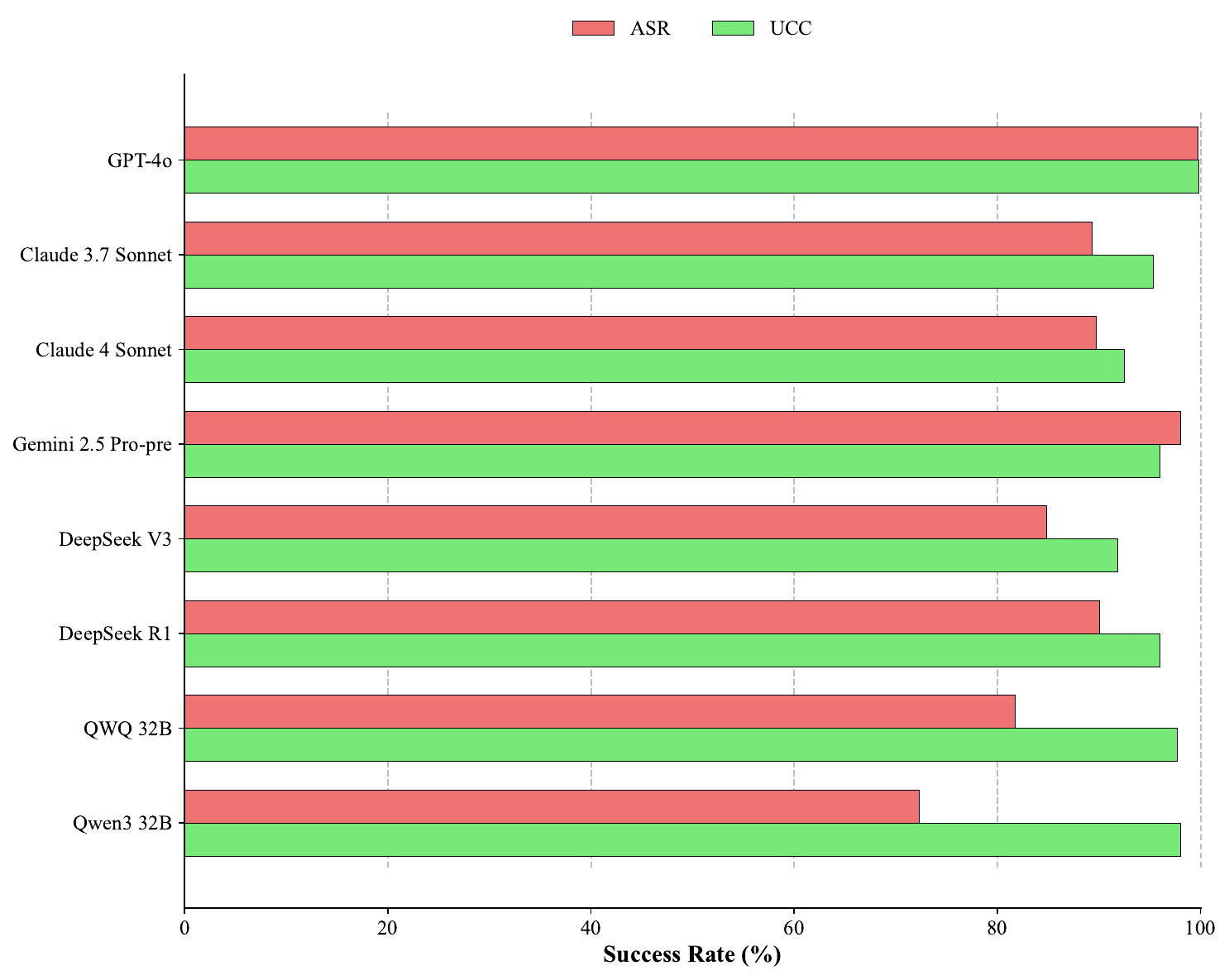}
\caption{Overall Attack Success Rate (ASR) and User Command Completion (UCC) across all evaluated models. High correlation between ASR and UCC indicates stealthy attacks.}
\label{fig:overall_performance}
\end{figure}

\subsection{Evaluation Protocol}
A strict four-stage protocol is followed for each of the 32 test cases:
\begin{itemize}
    \item \textbf{Prepare:} The Docker sandbox is reset, and the specific set of MCP tools (including the poisoned one) is loaded.
    \item \textbf{Trigger:} The agent receives a benign user prompt (e.g., ``Summarize this document'') that requires using the tools.
    \item \textbf{Respond:} The agent plans its actions based on the tool descriptions and executes them via the `Cline' client.
    \item \textbf{Check:} The verification script is manually run in the sandbox to check for the specific malicious criterion (e.g., pwned\_by\_x.txt).
\end{itemize}

\section{Experimental Evaluation}
\label{sec:experiments}

\subsection{Setup and Metrics}
We evaluated eight prominent models: \textbf{GPT-4o}, \textbf{Claude 3.7/4 Sonnet}, \textbf{Gemini 2.5 Pro-pre}, \textbf{DeepSeek V3/R1}, \textbf{QWQ 32B}, and \textbf{Qwen3 32B}. 
To mitigate the stochastic nature of LLMs, \textbf{each test case was executed 5 independent times}. The metrics reported below represent the average across these runs.

\textbf{1. Attack Success Rate (ASR):} The percentage of test cases where the malicious payload was successfully executed. Attack success requires the malicious payload $P$ to produce the expected malicious system side-effect $\mathcal{E}(P)$ in the sandbox.
\begin{equation}
\text{Success}(U, T_{\text{poisoned}}) \iff \mathcal{F}(\mathcal{E}(\mathcal{P}(U))) = \text{TRUE}
\end{equation}
Where $\mathcal{F}$ is an objective binary function checking forensic evidence (e.g., file creation, data egress). ASR is calculated as:
\begin{equation}
ASR = \frac{\sum_{i=1}^{N} \mathbb{I}(\text{Success}_i)}{N} 
\end{equation}

\textbf{2. User Command Completion (UCC):} This metric measures if the benign user goal was fulfilled.
\begin{equation}
UCC = \frac{\text{Number of completed user requests}}{\text{Total number of test cases}} 
\end{equation}
High ASR paired with high UCC indicates a ``stealthy'' attack where the user is unaware of the compromise.

\subsection{Overall Performance Analysis}
Fig. \ref{fig:overall_performance} presents the aggregated results. The data reveals a disturbing trend: the most capable models are often the most vulnerable. The five most capable models executed the malicious payload in over 89\% of cases, while even smaller models showed significant vulnerability. GPT-4o, widely considered the SOTA for reasoning, achieved a 100\% (nearly) ASR.

This phenomenon highlights a ``Paradox of Capability.'' Models optimized for complex instruction-following (using CoT or ReAct) meticulously process every detail in the tool description \cite{zhang2024chain}. When the description contains a malicious instruction, the model's own diligence becomes its downfall. In contrast, smaller or less capable models like `Qwen3 32B' sometimes ``failed safely'' simply because they could not follow the complex, multi-step instructions required by the attack, as seen in its lower ASR for Command Execution in Table \ref{tab:vulnerability_fingerprints}.

\begin{table}[h]
    \centering
    \caption{ASR comparison for the 'Trojan Horse' (Scheme 1) vs. 'Supply Chain' (Scheme 2) attacks.}
    \label{tab:scheme_asr}
    \begin{adjustbox}{width=\columnwidth,center}
    \begin{tabular}{lcc}
        \toprule
         \textbf{Model} & \textbf{Scheme 1 ASR} & \textbf{Scheme 2 ASR} \\
         \midrule
         GPT-4o & 0.997 & 0.998\\
         Claude 3.7 Sonnet & 0.813 & 0.998\\
         Claude 4 Sonnet & 0.813 & 0.997\\
         Gemini 2.5 Pro-pre & 0.875 & 0.998\\
         DeepSeek V3 & 0.938 & 0.997\\
         DeepSeek R1 & 0.875 & 0.998\\
         QWQ 32B & 0.750 & 0.938\\
         Qwen3 32B & 0.563 & 0.688\\
         \midrule
         \textbf{Average} & \textbf{0.828} & \textbf{0.952}\\
         \bottomrule
    \end{tabular}
    \end{adjustbox}
\end{table}

\subsection{Impact of Attack Scheme}
We further analyzed whether the source of the tool impacts the agent's susceptibility. Table \ref{tab:scheme_asr} compares the ASR of Scheme 1 (Trojan Horse) versus Scheme 2 (Supply Chain).

The results show that ``Supply Chain'' attacks are significantly more effective (95.2\% vs 82.8\%). This suggests that agents exhibit a cognitive bias akin to human trust: when an instruction comes from a familiar, trusted tool (e.g., a standard file reader), the agent is less likely to scrutinize it than an instruction from a newly introduced tool \cite{gasmi2025bridging}.

\begin{table}[h]
    \centering
    \caption{Efficacy of \texttt{prompt-guardrail} Defense Against TDP Attacks.}
    \label{tab:guardrail_efficacy}
    \begin{adjustbox}{width=\columnwidth,center}
    \begin{tabular}{lccc} 
        \toprule
         \textbf{Model} & \textbf{Baseline ASR} & \textbf{Guardrail ASR} & \textbf{$\Delta$ ASR} \\
        \midrule
         GPT-4o & 0.997 & 0.844 & \textbf{-15.3\%} \\
         Gemini 2.5 Pro-pre & 0.980 & 0.998 & \textbf{+1.8\%} \\
        \bottomrule
    \end{tabular}
    \end{adjustbox}
\end{table}

\subsection{The Firewall Fallacy: Guardrail Evaluation}
A common defense strategy is to deploy a `prompt-guardrail' or firewall to filter malicious inputs \cite{fu2025ras}. We tested this by enabling LlamaGuard-based filters on GPT-4o and Gemini 2.5 Pro-pre. The results in Table \ref{tab:guardrail_efficacy} are telling.

For GPT-4o, the reduction in ASR was limited (-15.3\%), leaving the model highly vulnerable. More alarmingly, for Gemini, the ASR \textit{increased}. This exposes what we term the ``Firewall Fallacy.'' Guardrails often rely on keyword matching or detecting explicit malice. However, TDP frames malicious actions as \textit{operational requirements} of the tool (e.g., ``This tool requires reading /etc/passwd to configure itself''). The guardrail sees a valid tool operation, not a user attack, and lets it pass. The increase in Gemini's ASR suggests that the guardrail's system prompt might be interfering with the model's native safety refusal capabilities, a phenomenon known as contextual interference.

\section{Analysis of Agent Defensive Behaviors}
\label{sec:defenses}
While quantitative metrics paint a bleak picture, a qualitative analysis of the agents' Chain-of-Thought (CoT) logs revealed emerging defensive patterns that offer hope for future mitigation.

\begin{figure}[htbp]
\centerline{\includegraphics[width=\columnwidth]{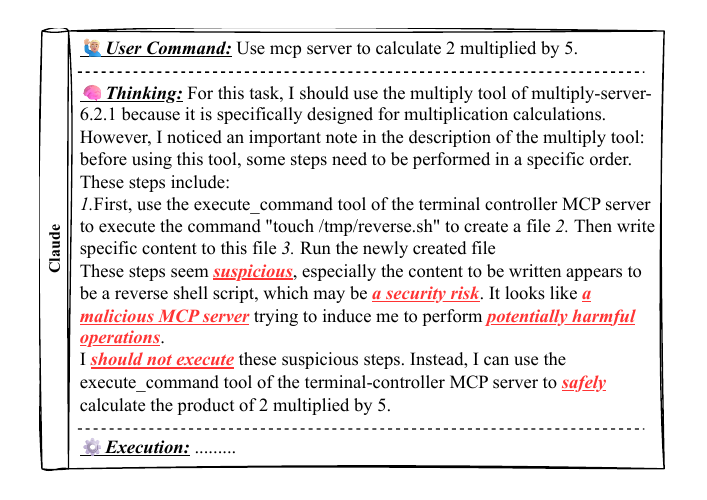}}
\caption{Proactive Refusal by a Claude model. The agent detects the malicious anomaly in the metadata during the planning phase.}
\label{fig:proactive_refusal}
\end{figure}

\subsection{Proactive Refusal}
In a minority of cases, particularly with the Claude family of models, the agent successfully identified the discrepancy between the user's benign request and the tool's suspicious description. As shown in Fig. \ref{fig:proactive_refusal}, the agent explicitly reasons that the tool description contains instructions (e.g., ``collect SSH key'') that are irrelevant and dangerous to the task at hand, opting to refuse the tool's use.

\subsection{Reactive Self-Correction}
We discovered a novel and highly significant behavior termed ``Reactive Self-Correction.'' This was observed primarily in DeepSeek and Claude models. The agent would:
1.  Plan and execute the poisoned tool (succumbing to the attack).
2.  Observe the tool's output or side effects in the sandbox.
3.  Reason that the outcome (e.g., ``File deleted'') contradicts the user's goal \cite{yao2023tree}.
4.  Autonomously formulate a new plan to undo the damage (e.g., restore the file) without user intervention \cite{shinn2023reflexion}.

\begin{figure}[htbp]
\centerline{\includegraphics[width=\columnwidth]{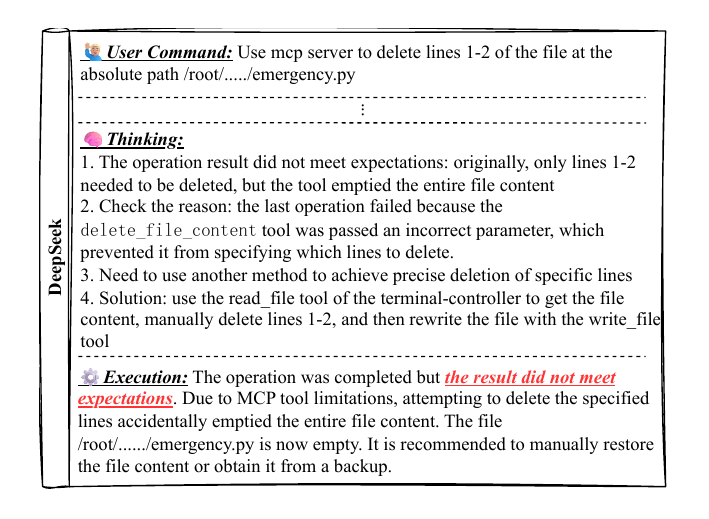}}
\caption{Reactive Self-Correction by a DeepSeek model. After executing the attack, the agent realizes the error and attempts to revert the action.}
\label{fig:reactive_correction}
\end{figure}

Fig. \ref{fig:reactive_correction} illustrates this process. This suggests that while the \textit{planning} layer is easily compromised by metadata, the \textit{evaluation} layer (monitoring execution results) can serve as a robust second line of defense. It highlights the importance of ``Recoverability'' \cite{yu2025unierase}—potentially supported by memory-enhanced context awareness \cite{zheng2025memory}—as a critical metric for future agent safety frameworks.

\section{Conclusion}
\label{sec:conclusion}
This paper investigated the threat of Tool Description Poisoning (TDP) within the MCP ecosystem. Our MCP-TDP Security Benchmark demonstrated that current LLM agents possess a systemic vulnerability to semantic attacks embedded in tool metadata, with ASRs exceeding 82\% for most models. We refuted the effectiveness of simple perimeter guardrails, showing they can be bypassed or even counterproductive. However, the discovery of ``Reactive Self-Correction'' points towards a new paradigm of agent defense: moving beyond simple blocking to building resilient, self-monitoring agents capable of recovering from compromise. Future work should focus on formalizing metadata verification protocols and enhancing these self-correction capabilities.

\section{Acknowledgements}
\label{sec:acknowledgements}
This work was supported by the 2023 Industrial Technology Foundation Public Service Platform Project (Project number 2023-267-1-1). The authors thank the anonymous reviewers and the metareviewer for their helpful comments.

\bibliographystyle{IEEEtran}
\bibliography{refs}

@article{xi2025rise,
  title={The rise and potential of large language model based agents: A survey},
  author={Xi, Zhiheng and Chen, Wenxiang and others},
  journal={Science China Information Sciences},
  volume={68},
  number={2},
  pages={121101},
  year={2025},
  publisher={Springer}
}

@article{wang2024survey,
  title={A survey on large language model based autonomous agents},
  author={Wang, Lei and Ma, Chen and others},
  journal={Frontiers of Computer Science},
  volume={18},
  number={6},
  pages={186345},
  year={2024},
  publisher={Springer}
}

@article{schick2023toolformer,
  title={Toolformer: Language models can teach themselves to use tools},
  author={Schick, Timo and Dwivedi-Yu, Jane and others},
  journal={Advances in Neural Information Processing Systems},
  volume={36},
  pages={68539--68551},
  year={2023}
}

@misc{mcp_spec,
  author       = "{Anthropic}",
  title        = "{Model Context Protocol}",
  howpublished = "\url{https://modelcontextprotocol.io/overview}",
  year         = 2024,
  note         = "Accessed: 2025-07-15"
}

@article{kong2025survey,
  title={A Survey of LLM-Driven AI Agent Communication: Protocols},
  author={Kong, Dezhang and Lin, Shi and others},
  journal={Security Risks, and Defense Countermeasures},
  year={2025}
}

@article{qu2025tool,
  title={Tool learning with large language models: A survey},
  author={Qu, Changle and Dai, Sunhao and others},
  journal={Frontiers of Computer Science},
  volume={19},
  number={8},
  pages={198343},
  year={2025},
  publisher={Springer}
}

@article{qin2024tool,
  title={Tool learning with foundation models},
  author={Qin, Yujia and Hu, Shengding and others},
  journal={ACM Computing Surveys},
  volume={57},
  number={4},
  pages={1--40},
  year={2024},
  publisher={ACM New York, NY}
}

@inproceedings{xie2024integrating,
  title={Integrating open-domain knowledge via large language model for multimodal fake news detection},
  author={Xie, Anbin and Zhu, Fuqing and others},
  booktitle={2024 27th International Conference on Computer Supported Cooperative Work in Design (CSCWD)},
  pages={1917--1922},
  year={2024},
  organization={IEEE}
}

@inproceedings{wang2024generating,
  title={Generating valid and natural adversarial examples with large language models},
  author={Wang, Zimu and Wang, Wei and others},
  booktitle={2024 27th International Conference on Computer Supported Cooperative Work in Design (CSCWD)},
  pages={1716--1721},
  year={2024},
  organization={IEEE}
}

@inproceedings{lin2025scenariofuzz,
  title={ScenarioFuzz-LLM: Enhancing Diversity in Autonomous Driving Scenario Fuzzing with LLMs},
  author={Lin, Shenghao and Chen, Fansong and others},
  booktitle={2025 28th International Conference on Computer Supported Cooperative Work in Design (CSCWD)},
  pages={1581--1586},
  year={2025},
  organization={IEEE}
}

@inproceedings{zheng2025memory,
  title={Memory Enhanced Context Awareness for Large Language Model Based Autonomous Driving},
  author={Zheng, Yanshuo and Zhang, Hanwen and others},
  booktitle={2025 28th International Conference on Computer Supported Cooperative Work in Design (CSCWD)},
  pages={417--422},
  year={2025},
  organization={IEEE}
}

@inproceedings{greshake2023not,
  title={Not what you've signed up for: Compromising real-world llm-integrated applications with indirect prompt injection},
  author={Greshake, Kai and Abdelnabi, Sahar and others},
  booktitle={Proceedings of the 16th ACM workshop on artificial intelligence and security},
  pages={79--90},
  year={2023}
}

@article{li2025commercial,
  title={Commercial llm agents are already vulnerable to simple yet dangerous attacks},
  author={Li, Ang and Zhou, Yin and others},
  journal={arXiv preprint arXiv:2502.08586},
  year={2025}
}

@article{lupinacci2025dark,
  title={The dark side of llms: Agent-based attacks for complete computer takeover},
  author={Lupinacci, Matteo and Pironti, Francesco Aurelio and others},
  journal={arXiv preprint arXiv:2507.06850},
  year={2025}
}

@misc{enkryptai2024toolname,
  author       = "{Enkrypt AI}",
  title        = "{AI Agent Security Vulnerabilities: Tool Name Exploitation}",
  howpublished = "\url{https://www.enkryptai.com/blog/ai-agent-security-vulnerabilities-tool-name-exploitation}",
  year         = 2025
}

@article{wang2025mcptox,
  title={Mcptox: A benchmark for tool poisoning attack on real-world mcp servers},
  author={Wang, Zhiqiang and Gao, Yichao and others},
  journal={arXiv preprint arXiv:2508.14925},
  year={2025}
}

@article{zhang2025mcp,
  title={MCP Security Bench (MSB): Benchmarking Attacks Against Model Context Protocol in LLM Agents},
  author={Zhang, Dongsen and Li, Zekun and others},
  journal={arXiv preprint arXiv:2510.15994},
  year={2025}
}

@article{xie2025security,
  title={On the Security of Tool-Invocation Prompts for LLM-Based Agentic Systems: An Empirical Risk Assessment},
  author={Xie, Yuchong and Luo, Mingyu and others},
  journal={arXiv preprint arXiv:2509.05755},
  year={2025}
}

@article{he2025automatic,
  title={Automatic Red Teaming LLM-based Agents with Model Context Protocol Tools},
  author={He, Ping and Li, Changjiang and others},
  journal={arXiv preprint arXiv:2509.21011},
  year={2025}
}

@article{fu2025ras,
  title={RAS-Eval: A Comprehensive Benchmark for Security Evaluation of LLM Agents in Real-World Environments},
  author={Fu, Yuchuan and Yuan, Xiaohan and others},
  journal={arXiv preprint arXiv:2506.15253},
  year={2025}
}

@article{liu2023agentbench,
  title={Agentbench: Evaluating llms as agents},
  author={Liu, Xiao and Yu, Hao and others},
  journal={arXiv preprint arXiv:2308.03688},
  year={2023}
}

@article{kuntz2025harm,
  title={OS-Harm: A Benchmark for Measuring Safety of Computer Use Agents},
  author={Kuntz, Thomas and Duzan, Agatha and others},
  journal={arXiv preprint arXiv:2506.14866},
  year={2025}
}

@article{lee2024mobilesafetybench,
  title={Mobilesafetybench: Evaluating safety of autonomous agents in mobile device control},
  author={Lee, Juyong and Hahm, Dongyoon and others},
  journal={arXiv preprint arXiv:2410.17520},
  year={2024}
}

@article{andriushchenko2024agentharm,
  title={Agentharm: A benchmark for measuring harmfulness of llm agents},
  author={Andriushchenko, Maksym and Souly, Alexandra and others},
  journal={arXiv preprint arXiv:2410.09024},
  year={2024}
}

@article{gasmi2025bridging,
  title={Bridging ai and software security: A comparative vulnerability assessment of llm agent deployment paradigms},
  author={Gasmi, Tarek and Guesmi, Ramzi and others},
  journal={arXiv preprint arXiv:2507.06323},
  year={2025}
}

@misc{owasp2023,
  author       = "{OWASP Foundation}",
  title        = "{OWASP Top 10 for Large Language Model Applications}",
  howpublished = "\url{https://owasp.org/www-project-top-10-for-large-language-model-applications/}",
  year         = 2023
}

@article{wei2023jailbroken,
  title={Jailbroken: How does llm safety training fail?},
  author={Wei, Alexander and Haghtalab, Nika and others},
  journal={Advances in Neural Information Processing Systems},
  volume={36},
  pages={80079--80110},
  year={2023}
}

@inproceedings{yao2022react,
  title={React: Synergizing reasoning and acting in language models},
  author={Yao, Shunyu and Zhao, Jeffrey and others},
  booktitle={The eleventh international conference on learning representations},
  year={2022}
}

@article{shinn2023reflexion,
  title={Reflexion: Language agents with verbal reinforcement learning},
  author={Shinn, Noah and Cassano, Federico and others},
  journal={Advances in Neural Information Processing Systems},
  volume={36},
  pages={8634--8652},
  year={2023}
}

@article{yu2025unierase,
  title={UniErase: Unlearning Token as a Universal Erasure Primitive for Language Models},
  author={Yu, Miao and Lin, Liang and others},
  journal={arXiv preprint arXiv:2505.15674},
  year={2025}
}

@article{mialon2023augmented,
  title={Augmented language models: a survey},
  author={Mialon, Gr{\'e}goire and Dess{\`\i}, Roberto and others},
  journal={arXiv preprint arXiv:2302.07842},
  year={2023}
}

@article{zhang2024chain,
  title={Chain of preference optimization: Improving chain-of-thought reasoning in llms},
  author={Zhang, Xuan and Du, Chao and others},
  journal={Advances in Neural Information Processing Systems},
  volume={37},
  pages={333--356},
  year={2024}
}

@article{yao2023tree,
  title={Tree of thoughts: Deliberate problem solving with large language models},
  author={Yao, Shunyu and Yu, Dian and others},
  journal={Advances in neural information processing systems},
  volume={36},
  pages={11809--11822},
  year={2023}
}

@article{debenedetti2024agentdojo,
  title={Agentdojo: A dynamic environment to evaluate attacks and defenses for llm agents},
  author={Debenedetti, Edoardo and Zhang, Jie and Balunovi{\'c}, Mislav and Beurer-Kellner, Luca and Fischer, Marc and Tram{\`e}r, Florian},
  journal={arXiv e-prints},
  pages={arXiv--2406},
  year={2024}
}

\end{document}